# 80-Channel WDM-MDM Transmission over 50-km Ring-Core Fiber Using a Compact OAM DEMUX and Modular 4×4 MIMO Equalization

**Junwei Zhang[1,†], Yuanhui Wen[1,†], Heyun Tan[2], Jie Liu[1,*], Lei Shen[3], Maochun Wang[4], Jiangbo Zhu[5], Changjian Guo[4], Yujie Chen[1], Zhaohui Li[1], and Siyuan Yu[1,5,*]**

1. School of Electronics and Information Engineering, State Key Laboratory of Optoelectronic Materials and Technologies, Sun Yat-sen University, Guangzhou 510275, China
2. School of Physics, State Key Laboratory of Optoelectronic Materials and Technologies, Sun Yat-sen University, Guangzhou 510275, China
3. Yangtze Optical Fiber and Cable Joint Stock Limited Company, State key Laboratory of Optical Fiber and Cable Manufacture technology No.9 Guanggu Avenue, Wuhan, Hubei, P.R.China
4. South China Academy of Advanced Optoelectronics, South China Normal University, Guangzhou 510006, China
5. Photonics Group, Merchant Venturers School of Engineering, University of Bristol, Bristol BS8 1UB, UK
*E-mail address: liujie47@mail.sysu.edu.cn; s.yu@bristol.ac.uk
†These authors contributed equally to this work.

**Abstract:** 8-OAM modes each carrying 10 wavelengths with 2.56-Tbit/s aggregated capacity and 10.24-bit/s/Hz spectral efficiency have been transmitted over 50-km specially designed ring-core fiber, using a compact OAM mode sorter and only modular 4×4 MIMO equalization.
**OCIS codes:** (060.2330) Fiber optics communications; (060.4230) Multiplexing; (060.1660) Coherent communications.

## 1. Introduction

Mode-division multiplexing (MDM) is a potential candidate for increasing the transmission capacity and spectral efficiency over a single fiber [1]. A main limitation to the scalability of many MDM schemes is that the complexity of multiple-input multiple-output (MIMO) equalization sharply increases with the number of fiber modes involved, as well as their differential mode delay (DMD) [2]. Schemes based on breaking the MIMO into smaller blocks have been proposed to improve the capacity of MDM systems without significantly increasing the MIMO-equalization complexity, using weakly coupled few-mode fibers (FMFs) [3-5], elliptical-core fibers (ECFs) [6,7], as well as the mode group multiplexing (MGM) systems [8,9].

MDM schemes based on ring-core fibers (RCFs) that can stably support linear-polarized (LP) or orbital angular momentum (OAM) mode-sets have been reported recently as an alternative approach for achieving low-complexity and high-capacity MDM transmission [10-14]. In RCFs with single radial mode index, the number of modes in each high-order mode group (MG) (MG order > 0) is fixed at 4, which can decrease the MIMO complexity by exploiting the orthogonality between the MGs to suppress inter-MG crosstalk. The coupling between adjacent RCF MGs reduces as the azimuthal mode order increases [15], making them more scalable towards the higher order mode space. In addition, RCF-based optical amplifiers can also theoretically provide more equalized gain for all guided signal modes because they have similar mode profiles [16, 17]. These characteristics make RCF based MDM systems highly attractive in future high-capacity optical fiber transmission systems. Several MDM/MGM transmissions over RCFs with length of 1-2 km [10,11], 10 km [12], 18 km [13], and 24 km [14] have been reported.

In this paper, we demonstrate, for the first time, WDM-MDM transmission over a 50-km specially designed low-loss, low inter-MG coupling RCF and using a novel compact OAM mode sorter with high modal resolution [18]. Only modular 4x4 MIMO equalization is required to handle the intra-MG mode coupling, while the MGs are sufficiently de-coupled from each other by the RCF's specially designed refractive index profile. We demonstrate this scheme by exploiting 8 OAM modes belonging to two MGs ($|l|$ =2 and 3). Each mode carries 10 WDM wavelengths at 25 GHz grid, modulated by 16-GBaud QPSK signals, achieving aggregated capacity of 2.56-Tbit/s, spectral efficiency (SE) of 10.24 bit/s/Hz, and to the best of our knowledge, a record SE-distance product of 512 bit/s/Hz-km for OAM-MDM transmission.

## 2. Experimental system

The cross section and refractive index profile of the RCF used in the experimental setup are shown in Fig. 1(a). The inner and outer radius of the RCF are 3.5 μm and 7.5 μm, respectively with a maximum core-cladding refractive index difference (Δ$n$) of 0.01. A ring-shape index notch, whose inner radius, outer radius and index difference from the fiber cladding are 4.5 μm, 5.5 μm, and 0.009, respectively, is placed within the ring-core area to minimise inter-MG coupling caused by micro-perturbations (micro-bending) [19]. The RCF supports four OAM MGs with azimuthal order from $|l|$=0 to $|l|$=3. Between OAM MGs of $|l|$=1 and 2 Δ$n_{eff}$ is 2.2×10$^{-3}$, while its value is 3.3×10$^{-3}$ between OAM MGs of |*l*|=2 and 3. The fiber preform is fabricated by a PCVD process and the attenuation for all guided modes is around 0.31dB/km.

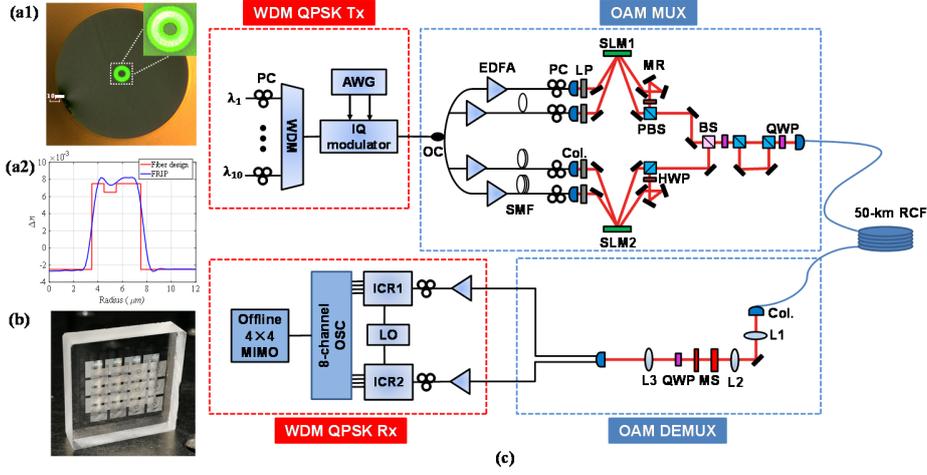

Fig. 1 (a1) The cross section and (a2) refractive index profile of the RCF; (b) an array of home-made OAM mode sorters consisting of unwrappers and phase correctors located on the both sides of a 5 mm-thick quartz plate; (c) Experiment setup. PC: polarization controller; OC: optical coupler; SMF: single-mode fiber; LP: linear polarizer; MR: mirror; QWP: quarter-wave plate; Col.: collimator; BS: beam splitter; MS: OAM Mode sorter DEMUX; ICR: integrated coherent receiver.

The compact OAM DEMUX device is based on the principle of a novel spiral coordinate transformation capable of high mode resolution OAM mode sorting [18]. As shown in Fig.1(b), the pair of transmissive diffractive optical elements (DOEs) implementing the spiral transformation are integrated onto both sides of a 5 mm-thick quartz plate. After this unitary mode transformation, OAM modes with different spiral wavefronts are mapped to plane wavefronts with different tilt angles, which can be directly separated in the focal plane of a Fourier transformation lens.

Fig. 1(c) illustrates the experimental setup of the OAM-MDM-WDM data transmission system. 10 optical carriers from narrow-linewidth tunable lasers with wavelengths ranging from 1549.8 nm to 1551.6 nm in a 0.2 nm/25-GHz grid are combined by a wavelength division multiplexer (WDM). Then the 10 WDM carriers are modulated by a 16-GBaud QPSK signals from an arbitrary waveform generator (AWG) through an I/Q modulator. A Bessel filter with a 3-dB bandwidth of 0.7×symbol rate is employed to eliminate the crosstalk between adjacent WDM channels, as shown in Fig. 2(c). Here we note that the adjacent WDM carriers are modulated with the same electrical signals due to the hardware limitation in our lab, which will have little impact on the system performance when guard bands exist between adjacent WDM channels. The sample rate of the digital-to-analog converter (DAC) is 64 GSa/s and the modulated electrical data sequence is pseudo-random binary sequence (PRBS) with pattern length of $2^{18}-1$.

The WDM signals are split into four branches which are separately delayed for data pattern decorrelation. After amplification by erbium-doped fiber amplifiers (EDFAs), collimation and linearly polarization, they are grouped into two pairs, each pair reflected by a phase-only spatial light modulator (SLM) for the conversion to OAM modes of *l* = +2 or +3. The pair of OAM modes off each SLM with one arm reflected by mirrors to invert the sign of OAM order are combined as orthogonally polarized channels to generate the four OAM modes of <+2, *X*>, <-2, *Y*>, <+3, *X*> and <-3, *Y*>. Here *X* and *Y* represent the horizontal and vertical polarization, respectively. Then the four modes are combined together, converted into circular polarizations and polarization multiplexed using a polarization beam splitter (PBS), an optical delay path and a polarization beam combiner (PBC), so that the four modes in each group of *l* = ±2 or ±3 with dual polarizations are created. The eight OAM modes are then multiplexed and converted into circular polarizations, and focused into the 50-km RCF.

After fiber transmission, all output modes from the fiber are collimated and then imaged to the plane of the DEMUX through a 4f beam expander that magnifies the OAM modes. Passing through the DEMX which employs an elliptical lens for the Fourier transform in the horizontal direction and beam focusing in the vertical direction, OAM modes with different *l* values are focused into Gaussian-like circular spots at different horizontal displacements proportional to *l*, which can be coupled into SMF arrays. For each MG (|*l*| = 2 or 3), the two Gaussian beams corresponding to +*l* and -*l* each contains two polarization multiplexed coherent signals, and are respectively coupled into two SMF-pigtailed dual-polarization integrated coherent optical receivers (ICRs). Then the eight output electrical waveforms (including the I/Q for each mode) from the ICRs are recorded by an 8-channel real-time oscilloscope (Teledyne LeCroy 10-36ZI) operated at 80-GSa/s for offline digital signal processing (DSP) including timing phase recovery, 4×4 MIMO equalizers based on conventional blind constant modulus algorithm (CMA), frequency offset compensation and carrier phase estimation, after which the BERs are finally evaluated. The measurement is repeated for both MGs and for each wavelength.

## 3. Results

Fig. 2(a) depicts measured power transfer matrix of the two multiplexed MGs over the entire system including the DEMUX. The crosstalk between mode groups $|l| = 2$ and $|l| = 3$ is measured to be about −10 dB. Fig. 2(b) shows the observed intensity profiles of multiplexed modes from groups $|l| = 2$ and $|l| = 3$. Fig. 2(c) shows the optical spectra of all the 10 wavelength channels. Constellations of the recovered 16-Gbaud QPSK signals after 50-km RCF transmission at their best OSNRs are shown in Fig. 2(d). Fig. 2(e) illustrates the measured BERs of all 8 OAM modes over the 10 WDM channels, with all 80 channels exceed the 20% soft-decision FEC threshold correcting BER of $2.4 \times 10^{-2}$. Here note that the BER evaluation for the two polarization signals of each OAM mode is performed together. Therefore, two BER dots are depicted for each OAM MG. Fig. 2(f) shows the convergent tap weights of the 16 FIR filters after 120 iterations of updating by using the CMA for both the two mode groups $|l| = 2$ and $|l| = 3$. The number of taps in each filter is set to 51, which is enough to cover the DMD within each MG. The DSP time consumption could be further reduced by using the frequency-domain MIMO equalization algorithm [4].

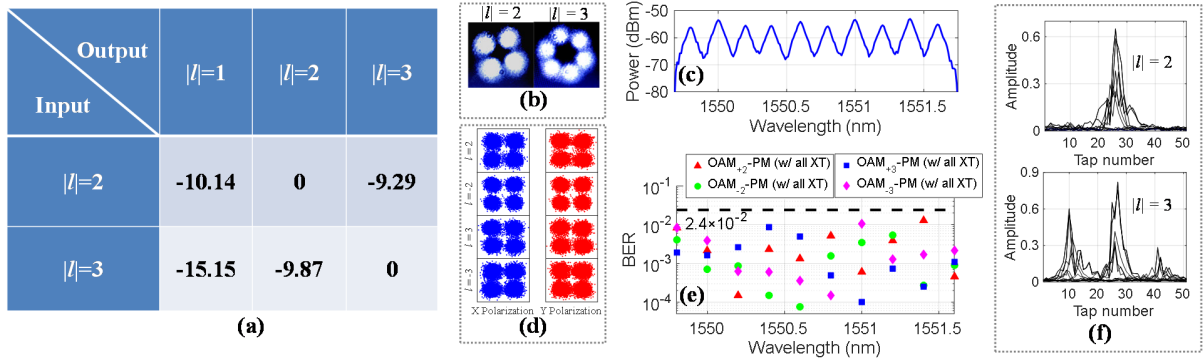

Fig. 2 (a) Measured power transfer matrix of the two multiplexed MGs over the entire system including the OAM DEMUX and 50-km RCF. (b) observed intensity profiles of multiplexed modes from groups $|l| = 2$ and 3 after 50-km RCF transmission; (c) optical spectra of all the 10 wavelength channels; (d) constellations of the received signals with the best measured BERs at wavelength of 1550 nm after 50-km RCF transmission; (e) measured BERs of all 80 channels after 50-km RCF transmission; (f) the absolute values of complex tap weights of the four FIR filters to equalize the four modes in mode group $|l| = 2$ and $|l| = 3$.

## 4. Conclusion

A MDM-WDM system with 8 OAM modes and 10 wavelengths has been demonstrated over a 50-km specially designed RCF with low fiber loss and low inter-MG crosstalk, employing a compact OAM mode DEMUX with high modal resolution and modular 4×4 MIMO equalizers. Successful transmission of 16-Gbaud QPSK signals with a total capacity of 2.56-Tbit/s and for each wavelength, a spectral efficiency (SE) of 10.24 bit/s/Hz has be achieved, with a SE-distance product of 512 bit/s/Hz-km.

## 5. Acknowledgement

This work was supported by the NSFC-Guangdong joint program grant No. U1701661, Natural Science Foundation of China under grant No. 61490715 and 61875233, and Local Innovative and Research Teams Project of Guangdong Pearl River Talents Program 2017BT01X121. The refractive index profile of ring core fiber has been patented by Sun Yat-sen University.